\documentstyle[bezier,epsfig,12pt,preprint,tighten,aps]{revtex}

\begin{document}

\draft

\title{\rightline{{\tt {March 1999}}}
\rightline{{\tt UM-P-99/07}}
\rightline{{\tt KIAS-P99023}}
\rightline{{\tt UDELHP-99/101}}
\ \\
Closing the Neutrinoless Double Beta Decay Window into Violations of the
Equivalence Principle and/or Lorentz Invariance}
\author{A. Halprin {\it (b,c)} and R.R. Volkas {\it (a)}}
\address{(a) School of Physics,
Research Centre for High Energy Physics,
The University of Melbourne,
Parkville 3052 Australia\\
(b) Department of Physics and Astronomy,
The University of Delaware\\
Newark, DE 19716, USA\\
(c) Korean Institute for Advanced Study\\
 207-43 Cheongryangri-dong, Dongdaemun-gu, Seoul 130-012, Korea\\
(halprin@udel.edu, r.volkas@physics.unimelb.edu.au)}

\maketitle

\begin{abstract}

We have examined Lorentz invariance and equivalence principle  
violations in the neutrino sector as manifested in neutrinoless double beta decay. 
We conclude that this rare decay
cannot provide a useful view of these exotic processes.   

\end{abstract}

\newpage

\section{Introduction}

During the past few years, neutrino oscillations have been used to 
explore exotic 
properties of neutrinos such as possible violations of Lorentz invariance
(VLI)\cite{cg1,cg2,ghklp} and/or violations of the equivalence principle
 (VEP)\cite{fly,hlp,imy,hl,g1,g2}.  Since neutrinolesss
 double beta decay has served as a window 
into neutrino masses for the last two decades \cite{hmpr,dkt,m}, it is natural to enquire 
if this rare decay can tell us anything about 
VLI and VEP processes.
 
We begin with the observation that the properties of neutrinos enter into
the neutrino exchange diagram for the 
neutrinoless double beta decay amplitude (without right handed currents) in the
form of the factor \cite{k}
\begin{equation}
A^{\nu}=(1-\gamma_5)P(1-\gamma_5),
\end{equation}
where P is a linear combination of the  propagators for each of the
Majorana neutrino fields  that constitute the $\nu_e$ field,
\begin{equation}
P=U_{ea}^2 P_a.
\end{equation}
The $U_{ea}$ are elements of the unitary matrix connecting mass and weak
eigenstate neutrinos.
In the absence of external fields, and in the absence of Lorentz
invariance violation, the Majorana fields have
definite masses, $m_a$, and all neutrinos have the same limiting velocity, $c$.
In that case the $ P_a$ are, of course, given by 
\begin{equation}
P_a=(\gamma^0E-\gamma^k p_k c +m_a c^2)^{-1}.
\end{equation}

\section{Modifications Due to Violation of Lorentz Invariance.}

In the Lorentz invariance violating scheme of Coleman and Glashow, the limiting 
velocity of each neutrino may be distinct, so that $c\rightarrow c_a$.  Our conclusions 
are unaltered by the introduction of distinct mass and velocity bases, and for the sake 
of clarity such a complication will be ignored.  To first order in $m_a$, the VLI modified
$A^{\nu}$ is then given by
\begin{equation}
A^{\nu}_{VLI} \simeq - \frac{2 U^2_{ea}(1 - \gamma_5) m_a c_a^2}{E^2- (p c_a)^2}
\end{equation}
where the chirality factors have been used to eliminate contributions from factors
of $E\gamma^0$ or $\gamma^k p_k$ in the numerator.

For neutrinoless double beta decay in nuclei we usually make the approximation of 
ignoring nuclear recoil, so the energy of the exchanged neutrino is set equal
to zero, that is $E=0$.  With this standard approximation,
\begin{equation}
A^{\nu}_{VLI} \simeq 2 U_{ea}^2 (1 - \gamma_5) m_a/p^2.
\end{equation}
Since this expession is independent of limiting velocities, we conclude that VLI cannot
enter into neutrinoless double beta decay in any significant way.

\section{Modifications Due to Violation of the Equivalence Principle.}

Following the formalism of Ref.\cite{imy}, 
the Dirac equation governing neutrino $a$ is modified by the presence of an external
gravitational field which couples to neutrinos with strength $f_a$ relative
to the usual universal Newtonian coupling. For the sake of clarity, 
we take the mass and gravitational coupling bases to be the same.  In the presence of a 
constant Newtonian potential, $\Phi$, the neutrino propagator becomes 
\begin{equation}
P_a = [(1+ f_a \Phi)E\gamma^0 -(1 - f_a\Phi)\gamma^k p_k + m_a]^{-1}
\end{equation}
where we have set the common limiting vacuum (i.e. $\Phi = 0$) velocity equal to 1.

To first order in $m_a$ we then have, making use of the chirality factors as before,
\begin{equation}
A^{\nu}_{VEP} \simeq - \frac{2 U_{ea}^2 (1 - \gamma_5) m_a}{(1 + f_a \Phi)^2 E^2 - (1 -
f_a\Phi)^2 p^2}.
\end{equation}
Making the zero recoil approximation as above and retaining only to first order in 
$\Phi$ for consistency, we then have
\begin{equation}
A^{\nu}_{VEP} \simeq 2 U_{ea}^2 (1 - \gamma_5)m_a(1 + 2f_a\Phi)/p^2.
\end{equation}

The expression above includes only the modification to the neutrino propagator 
due to the presence of $\Phi$. To this we must also add modifications to the W- boson,
quark and electron lines due to $\Phi$.  Assuming that only neutrinos 
have anomalous gravitational
couplings, restoration of gravitational gauge invariance in the limit that all
$f_a$ are equal guarantees 
that the final $\Phi$-dependent contribution depends only upon 
$\Delta f_a = f_a - f_0$, where $f_0 = 1$ in Einsteinian gravity.  The $\Phi$-dependent
contribution to the total neutrinoless double beta decay rate will then be
proportional to $U_{ea}^2 m_a \Delta f_a$.  Thus, the VEP effect is proportional to both
$m_a$ and $\Delta f_a$ and is therefore extremely small.

\section{Conclusions}

We have examined the modifications to the usual neutrino exchange
diagram for the neutrinoless double beta decay amplitude arising from violations 
of Lorentz invariance and/or the equivalence principle in the neutrino sector. 
We find that the VLI parameters disappear from the decay amplitude in the usual 
zero recoil approximation and that the VEP parameters enter the amplitude only
in combination with with neutrino mass factors. We therefore conclude that
neutrinolesss double beta decay cannot provide a significant window into 
VLI 
and VEP neutrino processes.  This result appears to contradict the 
conclusions of a recent paper \cite{kps}.

\acknowledgments{A.H. would like to thank the particle theory group at The University
of Melbourne and the Korean Institute for Advanced Study for 
their hospitality during a portion of this work. He also thanks C.N. Leung 
for discussions during the early stages of this work. This work was supported in part
by the US Department of Energy grant DE-FG02-84ER40163. R.R.V. was supported by the
Australian Research Council.}

\end{document}